\documentclass[a4paper,10pt,3p,final,pdftex]{elsarticle}

\biboptions{sort&compress}

\usepackage{amsmath}
\usepackage{amsfonts}
\usepackage{graphicx}
\usepackage{url}
\usepackage{color}
\usepackage{tikz}
\usepackage{multirow}

\newcommand{\matrx}[1]{\mathcal{#1}}
\renewcommand{\vec}[1]{\mathbf{#1}}
\newcommand{\figref}[1]{Fig.~\ref{fig:#1}}
\newcommand{\eqnref}[1]{\eqref{eqn:#1}}
\newcommand{\secref}[1]{Section~\ref{sec:#1}}

\DeclareFontFamily{U}{euc}{}
\DeclareFontShape{U}{euc}{m}{n}{<-6>eurm5<6-8>eurm7<8->eurm10}{}
\DeclareSymbolFont{AMSc}{U}{euc}{m}{n} 
\DeclareMathSymbol{\umu}{\mathord}{AMSc}{"16} 

\journal{Journal of Computational Physics}

\begin{document}
\begin{frontmatter}
\title{Perfectly Matched Layers in a Divergence Preserving ADI Scheme for Electromagnetics}
\author[psi,ethz]{C.~Kraus}

\author[psi]{A.~Adelmann\corref{cor1}}
\ead{andreas.adelmann@psi.ch}
\cortext[cor1]{Corresponding author}

\author[ethz]{P.~Arbenz}

\address[psi]{Paul Scherrer Institut, WBGB/132, 5232 Villigen, Switzerland}
\address[ethz]{ETH Zurich, Chair of Computational Science, 8092 Z\"urich, Switzerland}

\begin{abstract}
For numerical simulations of highly relativistic and transversely accelerated charged particles including radiation fast algorithms are needed. While the radiation in particle accelerators has wavelengths in the order of $100\;\umu\text{m}$ the computational domain has dimensions roughly 5 orders of magnitude larger resulting in very large mesh sizes. The particles are confined to a small area of this domain only. To resolve the smallest scales close to the particles subgrids are envisioned. For reasons of stability the alternating direction implicit (ADI) scheme by D. N. Smithe \emph{et al.} (J. Comput. Phys. 228 (2009) pp.7289-7299) for Maxwell equations has been adopted. At the boundary of the domain absorbing boundary conditions have to be employed to prevent reflection of the radiation. In this paper we show how the divergence preserving ADI scheme has to be formulated in perfectly matched layers (PML) and compare the performance in several scenarios. 
\end{abstract}
\end{frontmatter}

\section{Introduction}
For simulations of electromagnetic fields an implementation using finite differences suggests itself through its simplicity and low computational expense. The explicit Yee scheme~\cite{bib:yee1966} is very well studied and widely adopted. Although its explicit formulation is to its advantage when it comes to computing time per time step it is also its weakness in terms of numerical stability. The Courant-Friedrich-Lewy (CFL) condition limits the ratio of the length of a time step and the width of a mesh cell. If high spatial resolution is needed only in a small fraction of the computational domain, then this constraint causes computational overhead, since the smallest cell determines the largest time step to be used. To overcome this constraint on the time step an alternating direction implicit (ADI) scheme can be adopted~\cite{bib:namiki1999, bib:zheng2000}. In this scheme the time step is split in two substeps. For every component of the electromagnetic field the partial derivative, originating from the curl operator, with respect to one direction is updated alternately implicitly and explicitly while this choice is reversed for the other partial derivative. In 2D half of the partial derivatives vanish, thus simplifying the systems of equations. The implicit part of the equations for the electric and magnetic components are coupled in such a way that multiple independent systems of equations in only one dimension emerge. For the simplest case of a homogeneous linear medium and a perfect electric conductor surrounding it, the resulting matrices are symmetric tridiagonal with dominant diagonals. Hence the resulting system is stable and can be solved fast. This amounts to a Cholesky decomposition at the beginning of a simulation and forward and backward insertion in every time step and spatial direction.

In simulations of charged particles that include the induced electromagnetic fields it is important to calculate the current and the fields self-consistently. To achieve self-consistency the interpolation of the electromagnetic field a the position of a particle and the interpolation of the charge current at the vertices of the mesh have to preserve the charge conservation, $\nabla \cdot \vec{J} + \frac{\partial}{\partial t} \rho = 0$, in a discrete sense. Furthermore, the generated fields have to satisfy Maxwell's equations. The latter gives rise to the necessity of a preserved divergence of the fields (Gauss's law). Smithe \emph{et al.} ~\cite{bib:smithe} showed that the original ADI scheme for Maxwell's equations introduced in~\cite{bib:namiki1999} does not satisfy this requirement. After a thorough study of all second-order ADI schemes they derived the only ADI scheme for Maxwell's equations that does preserve the divergence.

For simulations of open domains absorbing boundary conditions with low reflection are used to confine the computational domain. Perfectly matched layers~\cite{bib:berenger1994} accomplish absorption by introducing an electric and a magnetic resistivity in layers at the boundary of the domain. The Maxwell's equations in these layers then become
\begin{align*}
 \varepsilon_0 \frac{\partial}{\partial t} \vec{E} + \matrx{\sigma} \vec{E} &= \nabla \times \vec{H}, \\
 \mu_0 \frac{\partial}{\partial t} \vec{H} + \matrx{\sigma^{*}} \vec{H} &= - \nabla \times \vec{E},
\end{align*}
where $\sigma$ and $\sigma^{*}$ are the electric and magnetic conductivities, respectively. The PML can be adapted to the ADI scheme~\cite{bib:liu2000}. To that end the conductivity term is split and added equally to both the new and the old time step. For large time steps this choice is not optimal as was shown in~\cite{bib:wang2003}. The reflections can be reduced if the conductivity term is not split in equal parts. Instead it is much more advantageous to add it in one substep and in one direction fully to the new time step while in the other direction it is added to the old time step. In the next substep this choice is then reversed. 

In the following section we will shortly introduce the original and the divergence preserving ADI scheme in the formulation of~\cite{bib:smithe}. We restrict our presentation to the 2D case. Then we will show how perfectly matched layers are formulated using the divergence preserving scheme. To confirm the validity of the resulting formulation we conduct similar numerical experiments as in~\cite{bib:wang2003} and compare the results with the ones produced with the Yee scheme and the original ADI scheme in \secref{numexp}.
\section{Alternating Direction Implicit Schemes}
Using the four operators that were elaborated in~\cite{bib:smithe} the original ADI scheme for Maxwell's equations reads
\begin{equation*}
\left(1 + \frac{c\,\Delta t}{2} \matrx{P}\right)^{-1} \cdot \left(1 - \frac{c\,\Delta t}{2} \matrx{M}\right) \cdot \vec{V}^{n+1} = \left(1-\frac{c\,\Delta t}{2} \matrx{P}\right)^{-1} \cdot \left(1+\frac{c\,\Delta t}{2}\matrx{M}\right)\; \cdot \vec{V}^{n} + c\,\Delta t\, \vec{S}^{n+1/2},
\label{eqn:nondivpresADI}
\end{equation*}
where in 2D
\begin{align*}
\vec{V}^n &= 
\left(
\begin{array}{c}
E_x^n\\
E_y^n\\
Z_0\, H_z^n 
\end{array}
\right), 
&\matrx{P} \cdot \vec{V}^n =\;\;\;& 
\left(
\begin{array}{c}
Z_0 \matrx{D}^b_y\, H_z / \Delta y\\
0 \\
\matrx{D}^f_y E_x^n / \Delta y
\end{array}
\right), \\
\vec{S}^{n+1/2} &= -
\left(
\begin{array}{c}
Z_0\, J_x^{n+1/2}\\
Z_0\, J_y^{n+1/2}\\
0
\end{array}
\right), 
&\matrx{M}\cdot \vec{V}^n = 
-&\left(
\begin{array}{c}
0\\
Z_0 \matrx{D}^b_x\, H_z / \Delta x\\
\matrx{D}^f_x\, E_y^n / \Delta x
\end{array}
\right).
\end{align*}
Here, $Z_0$ is the impedance and $c$ the speed of light in vacuum, $J$ the current density and $\matrx{D}_x^f$, $\matrx{D}_x^b$ the forward and backward difference operators in \mbox{x-direction}, respectively. Their action on $E_y$ is described by 
\begin{align*}
\matrx{D}_x^f E_{y;\,i,\,j} &= E_{y;\,i+1,\,j} - E_{y;\,i,\,j}\;\text{ and }\\
\matrx{D}_x^b E_{y;\,i,\,j} &= E_{y;\,i,\,j} - E_{y;\,i-i,\,j}.
\end{align*} 
The operators $\matrx{D}_y^f$ and $\matrx{D}_y^b$ are the corresponding operators in \mbox{y-direction}.  In this notation the curl operator is $(\matrx{M} + \matrx{P})$.

The divergence preserving ADI scheme is built up using the same set of operators but applied in a different order:
\begin{equation}
\label{eqn:divpresADI}
\left(1 - \frac{c\,\Delta t}{2} \matrx{P}\right) \cdot \left(1 + \frac{c\,\Delta t}{2} \matrx{M}\right)^{-1} \cdot \vec{V}^{n+1} = \left(1+\frac{c\,\Delta t}{2} \matrx{P}\right) \cdot \left(1-\frac{c\,\Delta t}{2}\matrx{M}\right)^{-1}\; \cdot \vec{V}^{n} + c\,\Delta t\, \vec{S}^{n+1/2}.
\end{equation}
{\bf Remark:} In 3D the vectors $\vec{V}$, $\vec{S}$ and the operators $\matrx{M}$, $\matrx{P}$ have to be adapted:
\begin{align*}
\vec{V}^n &= 
\left(
\begin{array}{c}
E_x^n\\
E_y^n\\
E_z^n\\
Z_0\, H_x^n \\
Z_0\, H_y^n \\
Z_0\, H_z^n 
\end{array}
\right), 
&\matrx{P} \cdot \vec{V}^n =\;\;\;& 
\left(
\begin{array}{c}
Z_0 \matrx{D}^b_y\, H_z / \Delta y\\
Z_0 \matrx{D}^b_z\, H_x / \Delta z\\
Z_0 \matrx{D}^b_x\, H_y / \Delta x\\
\matrx{D}^f_z\, E_y^n / \Delta z \\
\matrx{D}^f_x\, E_z^n / \Delta x \\
\matrx{D}^f_y\, E_x^n / \Delta y 
\end{array}
\right), \\
\vec{S}^{n+1/2} &= -
\left(
\begin{array}{c}
Z_0\, J_x^{n+1/2}\\
Z_0\, J_y^{n+1/2}\\
Z_0\, J_z^{n+1/2}\\
0\\
0\\
0
\end{array}
\right), 
&\matrx{M}\cdot \vec{V}^n = 
-&\left(
\begin{array}{c}
Z_0 \matrx{D}^b_z\, H_y / \Delta z\\
Z_0 \matrx{D}^b_x\, H_z / \Delta x\\
Z_0 \matrx{D}^b_y\, H_x / \Delta y\\
\matrx{D}^f_y\, E_z^n / \Delta y \\
\matrx{D}^f_z\, E_x^n / \Delta z \\
\matrx{D}^f_x\, E_y^n / \Delta x
\end{array}
\right).
\end{align*}

\section{Perfectly Matched Layers}
The absorption of electromagnetic waves in PML is accomplished by introducing layers of artificial media. Each medium is described by an electric and a magnetic conductivity. The electric conductivity, $\sigma$, causes an exponential reduction of the amplitudes of incoming waves as $e^{-\sigma_x Z_0 x}$. The magnetic conductivity, $\sigma^{*}$, on the other hand is chosen such that the impedance of the medium matches the one of vacuum. To reduce the numerical reflections caused by abrupt changes in the conductivities, media with increasing conductivities towards the boundary of the domain are applied. The boundary of the domain is then terminated by a perfect electric conductor (PEC). The total theoretical reflection of a plane wave is described by $R(\varphi) = e^{-2 Z_0 \cos \varphi \int_0^\delta \sigma(r) dr}$ where $\varphi$ is the angle between the normal of the boundary and the direction of propagation of the incident wave and $\delta$ is the width of the PML region~\cite{bib:berenger1994}. 

In 2D the electric conductivity is split in two components, $\sigma_x$ and $\sigma_y$, which are associated with the y- and x-component of the electric field respectively. Also the magnetic field is split in two subcomponents, $H_{zx}$ and $H_{zy}$ (in 3D we have an additional four components of the magnetic field $H_{xy}$, $H_{xz}$, $H_{yx}$, $H_{yz}$ and two components of the electric field $E_{zx}$, $E_{zy}$ while the two existing components are split in four subcomponents $E_{xy}$, $E_{xz}$, $E_{yx}$, $E_{yz}$). Finally, the magnetic conductivity is split in components, $\sigma_x^{*}$ and $\sigma_y^{*}$, associated with $H_{zx}$ and $H_{zy}$, respectively.
\begin{figure}[h]
  \begin{center}
    \includegraphics[scale=1.0, angle=0]{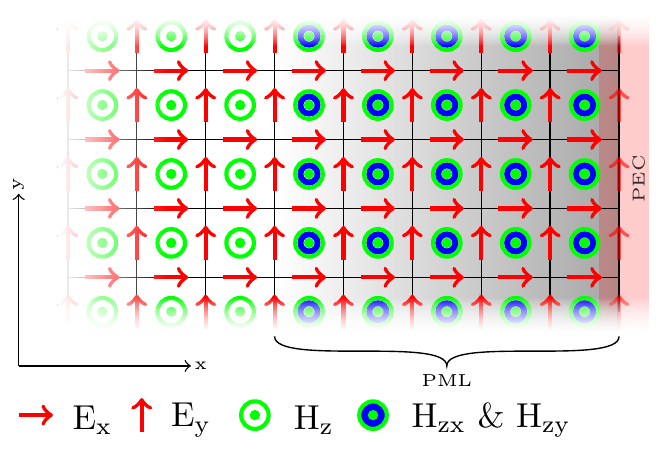}
    \caption{Example of a domain terminated by five layers of perfectly matched media. The increase of $\sigma_y$ and $\sigma_y^{*}$ towards the boundary of the domain is visualized by the shading.}
  \end{center}
\end{figure}

In our approach we add the conductivities $\sigma_x$, $\sigma_y$, $\sigma_x^{*}$ and~$\sigma_y^{*}$ to the four operators in \eqnref{divpresADI}.
The divergence preserving ADI scheme including PML then reads
\begin{equation} \label{eqn:adipml}
\left(\alpha^{3} - \frac{c\,\Delta t}{2} \matrx{P}\right) \cdot \left(\alpha^{4} + \frac{c\,\Delta t}{2} \matrx{M}\right)^{-1} \cdot \vec{V}^{n+1} = \left(\alpha^{2}+\frac{c\,\Delta t}{2} \matrx{P}\right) \cdot \left(\alpha^{1}-\frac{c\,\Delta t}{2}\matrx{M}\right)^{-1}\; \cdot \vec{V}^{n} + c\,\Delta t\, \vec{S}^{n+1/2},
\end{equation}
where
\begin{align*}
\vec{V}^n &= 
\left(
\begin{array}{c}
E_x^n\\
E_y^n\\
Z_0\, H_{zx}^n \\
Z_0\, H_{zy}^n
\end{array}
\right), 
&\matrx{P} \cdot \vec{V}^n =\;\;\;& 
\left(
\begin{array}{c}
Z_0 \matrx{D}^b_y\, \left(H_{zx}^n + H_{zy}^n\right) / \Delta y\\
0 \\
0 \\
\matrx{D}^f_y E_x^n / \Delta y
\end{array}
\right), 
\\
\vec{S}^{n+1/2} &= -
\left(
\begin{array}{c}
Z_0\, J_x^{n+1/2}\\
Z_0\, J_y^{n+1/2}\\
0 \\
0
\end{array}
\right), 
&\matrx{M}\cdot \vec{V}^n = 
-&\left(
\begin{array}{c}
0\\
Z_0 \matrx{D}^b_x\, \left(H_{zx}^n + H_{zy}^n\right) / \Delta x\\
\matrx{D}^f_x\, E_y^n / \Delta x\\
0
\end{array}
\right),
\end{align*}
and, for $\tau = \{1,\,2,\,3,\,4\}$,
\begin{align*}
\alpha^{\tau} &= 
\left(
\begin{array}{c}
1 + \lambda_{x}^{e,\,\tau} \frac{\Delta t \;\sigma_{y}}{2\,\varepsilon_0} \\
1 + \lambda_{y}^{e,\,\tau} \frac{\Delta t \;\sigma_{x}}{2\,\varepsilon_0} \\
1 + \lambda_{x}^{h,\,\tau} \frac{\Delta t \;\sigma^{*}_{x}}{2\,\mu_0} \\
1 + \lambda_{y}^{h,\,\tau} \frac{\Delta t \;\sigma^{*}_{y}}{2\,\mu_0} \\
\end{array}
\right), 
\quad\text{with } 
\begin{array}{cc}
\medskip
\lambda_\xi^{s,\, 2} = \lambda_\xi^{s,\,1} - 1,\quad &0 \le \lambda_\xi^{s,1} \le 1,\\
\lambda_\xi^{s,\, 4} = \lambda_\xi^{s,\,3} - 1,\quad & 0 \le \lambda_\xi^{s,3} \le 1.
\end{array}
\end{align*}
With $\lambda_{\xi}^{s,\,1}$ and $\lambda_{\xi}^{s,\,3}$ one can control whether the conductivity terms are added fully to quantities at the new time step (+1) or to the quantities at the old time step (0) or a linear combination thereof. 

The first substep of our new scheme inside the PML reads
\begin{align}
  E_y^{n+1/2} &= \alpha_y^{e,\,2} \left(\alpha_y^{e,\,1} - \frac{\eta_x^2}{4}\matrx{D}^b_x\frac{1}{\alpha_x^{h,\,1}} \matrx{D}^f_x \right)^{-1} \left(E_y^{n} - \frac{\eta_x\,Z_0}{2}\matrx{D}^b_x\left(\frac{1}{\alpha_x^{h,\,1}} H_{zx}^{n} + \frac{1}{\alpha_y^{h,\,1}} H_{zy}^{n}\right)\right), \label{eqn:implicitey}\\
  H_{zx}^{n+1/2} &= \frac{\alpha_x^{h,\,2}}{\alpha_x^{h,\,1}} \left(H_{zx}^{n} - \frac{\eta_x}{2\,Z_0} \matrx{D}^f_x \frac{1}{\alpha_y^{e,\,2}} E_y^{n+1/2}\right), \nonumber\\
  H_{zy}^{n+1/2} &= \frac{\alpha_y^{h,\,2}}{\alpha_y^{h,\,1}} H_{zy}^{n} + \frac{\eta_y}{2\,Z_0} \matrx{D}^f_y \frac{1}{\alpha_x^{e,\,1}} E_x^{n}, \nonumber\\
  E_x^{n+1/2} &= \frac{\alpha_x^{e,\,2}}{\alpha_x^{e,\,1}} E_x^{n} + \frac{\eta_y\,Z_0}{2}\matrx{D}^b_y \left(\frac{1}{\alpha_x^{h,\,2}} H_{zx}^{n+1/2} + \frac{1}{\alpha_y^{h,\,1}} H_{zy}^{n}\right). \nonumber
\end{align}
The second substep is then given by
\begin{multline}
  E_x^{n+1} =  \alpha_x^{e,\,4} \left(\alpha_x^{e,\,3} - \frac{\eta_y^2}{4} \matrx{D}^b_y \frac{1}{\alpha_y^{h,\,3}} \matrx{D}^f_y\right)^{-1}\\
  \left(E_x^{n+1/2} + \frac{\eta_y\,Z_0}{2} \matrx{D}^b_y\left(\frac{1}{\alpha_x^{h,\,3}} H_{zx}^{n+1/2} + \frac{1}{\alpha_y^{h,\,3}} H_{zy}^{n+1/2}\right) - c\Delta t Z_0 J_x^{n+1/2}\right),\label{eqn:implicitex}
\end{multline}
\begin{align*}  
H_{zx}^{n+1} &= \frac{\alpha_x^{h,\,4}}{\alpha_x^{h,\,3}} H_{zx}^{n+1/2} - \frac{\eta_x}{2\,Z_0} \matrx{D}^f_x \frac{1}{\alpha_y^{e,\,3}} \left(E_y^{n+1/2} - c\Delta t Z_0 J_y^{n+1/2}\right), \nonumber\\
H_{zy}^{n+1} &= \frac{\alpha_y^{h,\,4}}{\alpha_y^{h,\,3}} \left(H_{zy}^{n+1/2} + \frac{\eta_y}{2} \matrx{D}^f_y \frac{1}{\alpha_x^{e,\,4}} E_x^{n+1}\right),\nonumber \\
E_y^{n+1} &= \frac{\alpha_y^{e,\,4}}{\alpha_y^{e,\,3}} \left(E_y^{n+1/2} - c\Delta t Z_0 J_y^{n+1/2}\right) - \frac{\eta_x\,Z_0}{2} \matrx{D}^b_x \left(\frac{1}{\alpha_x^{h,\,3}} H_{zx}^{n+1/2} + \frac{1}{\alpha_y^{h,\,4}} H_{zy}^{n+1} \right). \nonumber
\end{align*}
As one can see from the inverted operators in \eqnref{implicitey} and \eqnref{implicitex} the systems that have to be solved are only in one dimension and the resulting matrices are tridiagonal. This is due to the fact that $\matrx{P}$ couples $E_x$ and $H_{zy}$ only and the operator $\matrx{M}$ couples $E_y$ and $H_{zx}$. In 3D a one-dimensional system has to be solved for every component of the electric field.

In~\cite{bib:wang2003} it is shown that for the non-divergence-preserving scheme one gets the best result if in one substep the conductivity is added fully to the quantities at the new time step of the pair $(E_x,\,H_{zy})$ while it is added fully to the quantities at the old time step of the other pair, $(E_y,\,H_{zx})$. In the second substep it is the other way round, for $(E_x,\,H_{zy})$ it is added fully explicitly and to $(E_y,\,H_{zx})$ it is added fully implicitly. For the divergence preserving scheme we find the combination 
\begin{center}
\begin{tabular}{l l}
$\lambda_x^{e,\,1} = \lambda_y^{h,\,1} = 0, \qquad$ & $\lambda_y^{e,\,1} = \lambda_x^{h,\,1} = 1,$\\
$\lambda_x^{e,\,2} = \lambda_y^{h,\,2} = -1$,       & $\lambda_y^{e,\,2} = \lambda_x^{h,\,2} = 0,$\\
$\lambda_x^{e,\,3} = \lambda_y^{h,\,3} = 1$,        & $\lambda_y^{e,\,3} = \lambda_x^{h,\,3} = 0,$\\
$\lambda_x^{e,\,4} = \lambda_y^{h,\,4} = 0$,        & $\lambda_y^{e,\,4} = \lambda_x^{h,\,4} = -1,$\\
\end{tabular}
\end{center}
 to yield the lowest reflections. This choice for $\lambda_{\xi}^{s,\,\tau}$ adds the conductivity term to the new time step if the operator in which it appears is applied implicitly and adds it to the old time step if the operator is applied explicitly. In 3D the set of $\alpha$'s is augmented by the $\alpha$'s of the additional components of the electromagnetic field. The rule how to choose the $\lambda$'s is however not altered.
\section{Numerical Experiments}
\label{sec:numexp}
\begin{figure}
  \begin{center}
    \includegraphics[scale=0.7,angle=0]{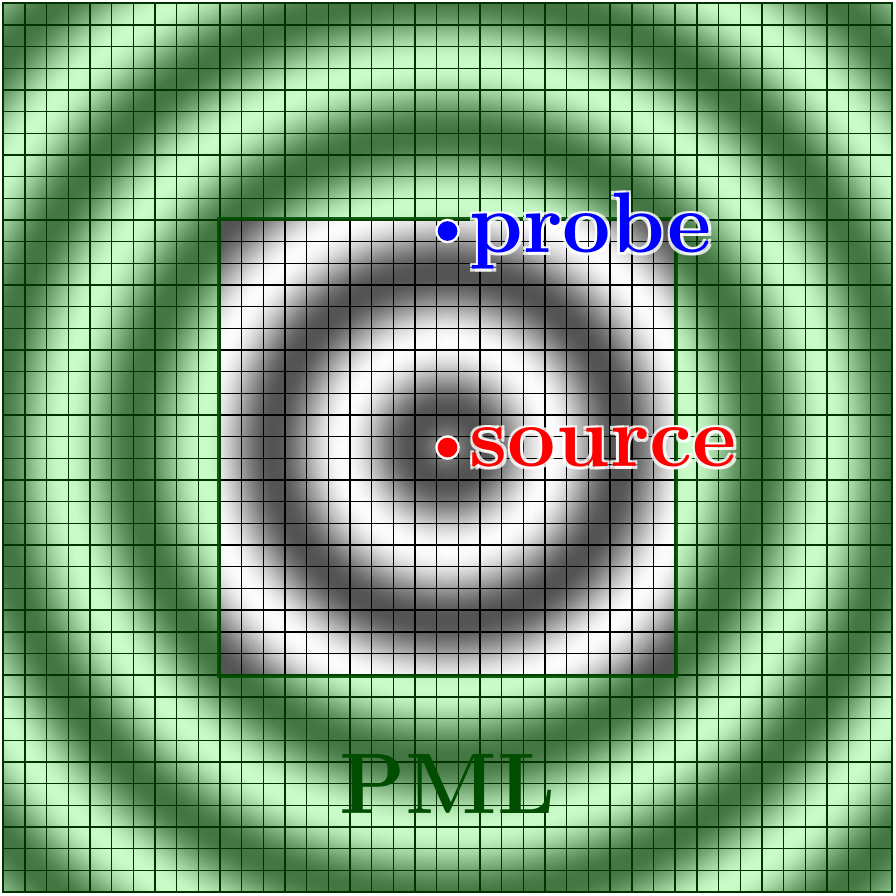}
    \caption{Setup of the numerical experiment with a soft source in the center of the domain, a probe near the interface to the PML  and 10 Layers of PML on the boundary of the domain.}
    \label{fig:experimentsetup}
  \end{center}
\end{figure}
To validate the performance of PML in the divergence preserving ADI scheme a test domain with 21 cells in each direction is excited by a source at the center, see \figref{experimentsetup}. The domain is enclosed by 10 perfectly matched layers on each boundary with 
\[
\sigma(r) = \frac{\sigma_m r^4}{\delta^4},
\]
where $r$ is the distance to the interface between vacuum and PML. $\sigma_m$ is chosen such that the theoretical reflection at an incident angle of 0 is $e^{-16}$~\cite[Ch. 7.6]{bib:taflove}. The source is a sine wave with frequency $3.175~\text{GHz}$ modulated by a differentiated Gaussian pulse with half bandwidth of $3.172~\text{GHz}$~\cite{bib:wang2003}, 
\[
\tilde{H}_z(x_0,\,y_0) = H_z(x_0,\,y_0) - \frac{t-t_0}{\tau^2} \; \sin(\omega \cdot t) \; e^{-\frac{(t-t_0)^2}{2\tau^2}},
\]
with $\omega = 2\pi \cdot 3.175\cdot10^9$, $t_0 = 0$, $\tau = \frac{\kappa_0 2\pi}{\omega}$ and $\kappa_0 e^{-\frac{\kappa_0^2-1}{2}} = 0.01$. The simulation is run for $1.5\text{ ns}$ or $\sim4.75$ periods of the central frequency.

A probe to sample the magnetic field is placed 10 cells away from the source, next to the interface between vacuum and PML. As a reference signal the magnetic field at a single point in a larger domain is recorded. This reference domain has 411 cells in each direction and is terminated by PEC boundary conditions but has no PML. The source is also placed at the center of the domain and the location of the probe is 10 cells away. The size of the reference domain ensures that the backscattered waves do not reach the probe during the simulation. The relative difference of the readings of the two setups,
\[
R = 20\,\log_{10}\left(\frac{|H_z(t) - H_{z;\;\text{ref}}(t)|}{\underset{t}{\max}|H_{z;\;\text{ref}}(t)|}\right),
\]
 versus time is plotted. The length of the time step in the simulation relative to the maximal time step for the Yee scheme is characterized by the Courant number $\chi = \Delta t / \Delta t_\text{max}$ where $\Delta t_\text{max} = \Delta x / (\sqrt{2} v_p)$ and $v_p$ is the phase velocity.
\subsection{Small Time Steps}
\begin{figure}
  \begin{center}
    \begin{tabular}{cc}
      \includegraphics[scale=0.35, angle=0]{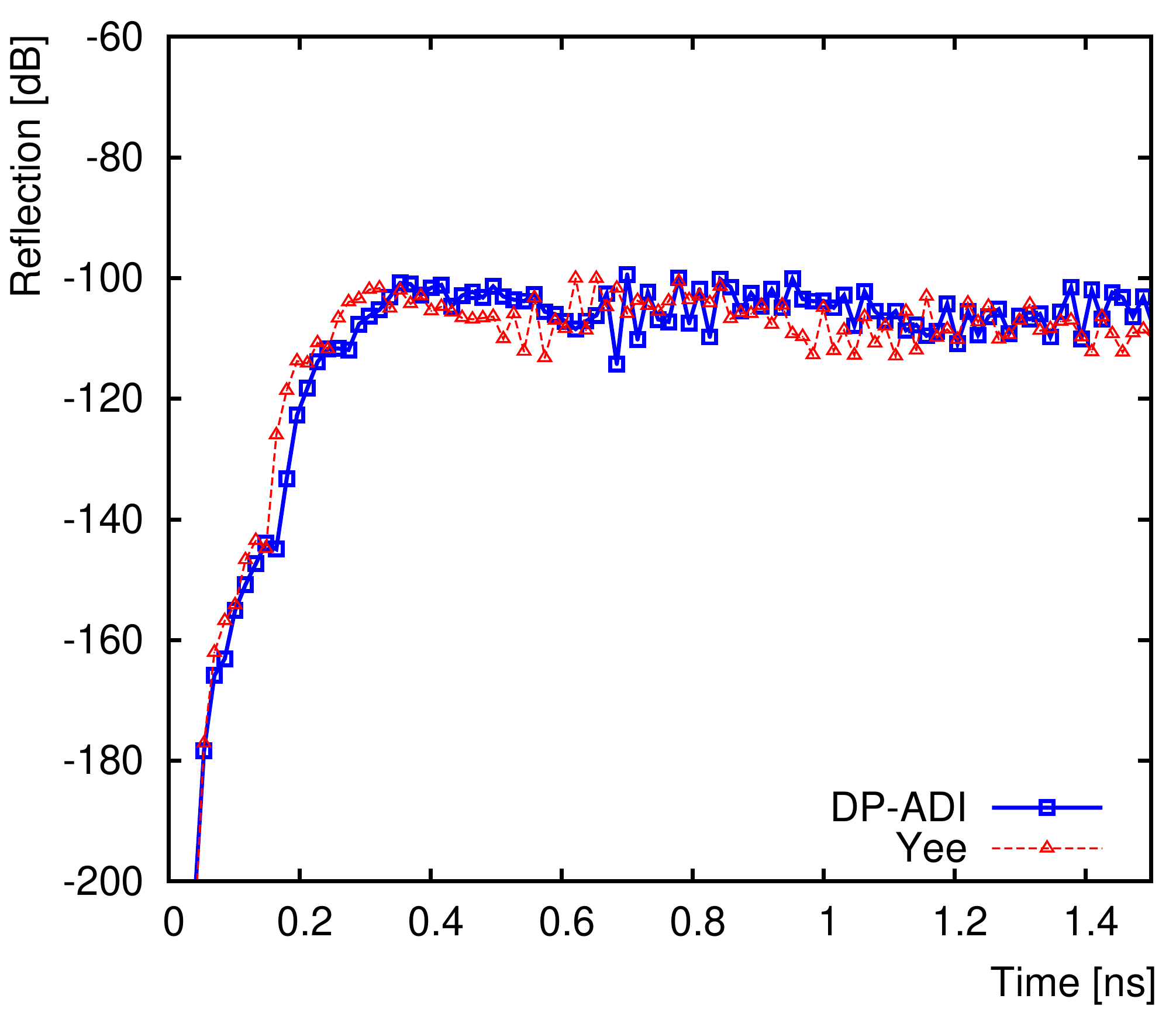} &
      \includegraphics[scale=0.35, angle=0]{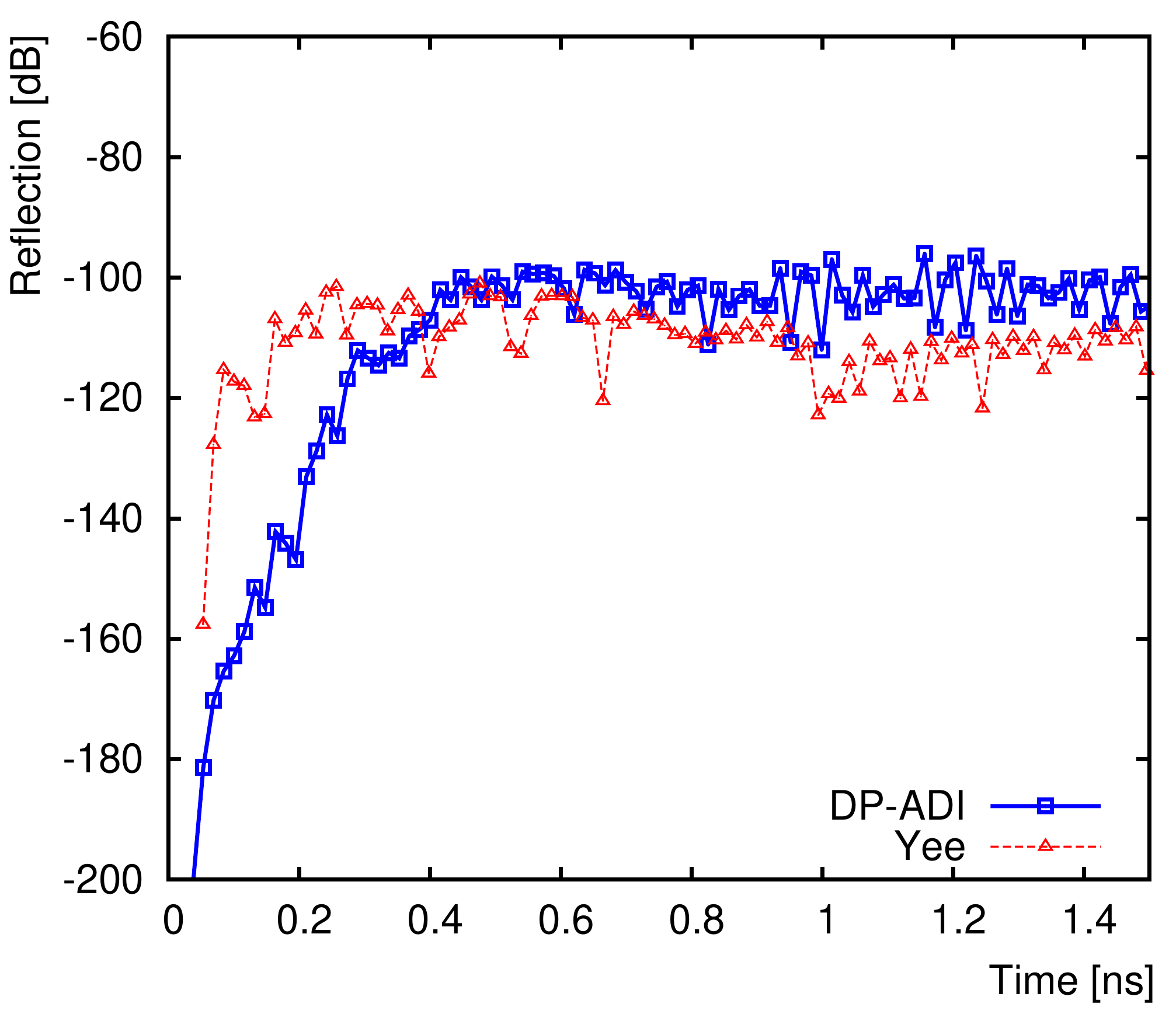}
    \end{tabular}
    \caption{Comparing the reflection at layers of perfectly matched media of the Divergence Preserving ADI and Yee scheme for $\chi = 0.5$ (left) and $\chi = 1$ (right). To reduce the noise the data was smoothed using a rolling average. For the case $\chi=0.5$ a length of 10 data points is used and only every 10$^\text{th}$ point is drawn. For the case $\chi=1$ only half of this length is used for the window and every 5$^\text{th}$ point is drawn}
    \label{fig:PML_reflection_smithe}
  \end{center}
\end{figure}

For a Courant number $\chi = 0.5$ the reflections for the divergence preserving ADI scheme and the Yee scheme are depicted in \figref{PML_reflection_smithe} on the left. On the right side data of a simulation with $\chi=1$, i.e.\ the largest time step feasible with the Yee scheme, are depicted. It is clearly seen that for small time steps the PML for the divergence preserving ADI scheme performs equally well as the Yee scheme. For $\chi = 1$ the Yee scheme performs slightly better but the results of both schemes are still below $-90~\text{dB}$.
\subsection{Large Time Steps}
\begin{figure}
  \begin{center}
    \begin{tabular}{cc}
      \includegraphics[scale=0.35, angle=0]{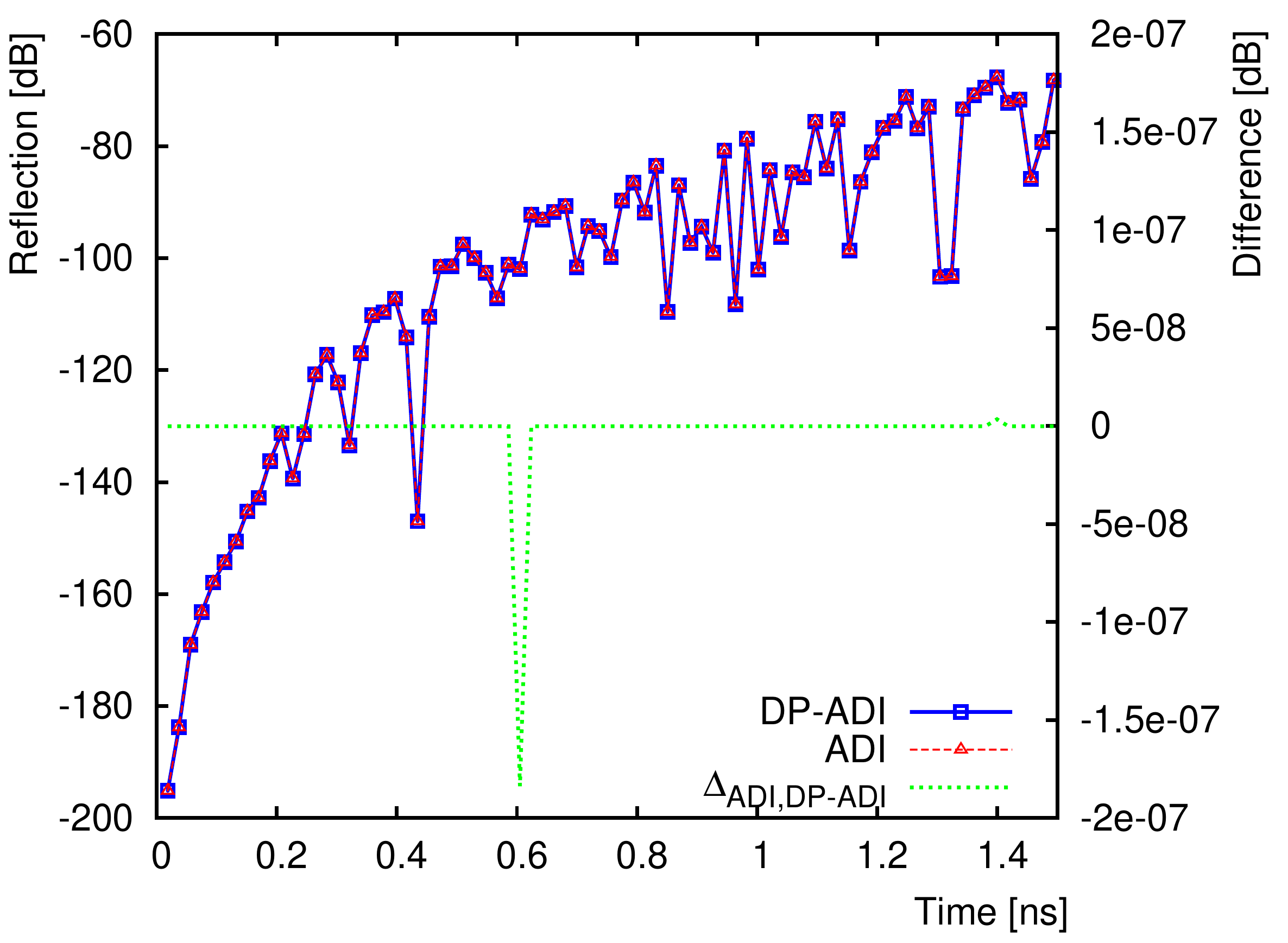}&
      \includegraphics[scale=0.35, angle=0]{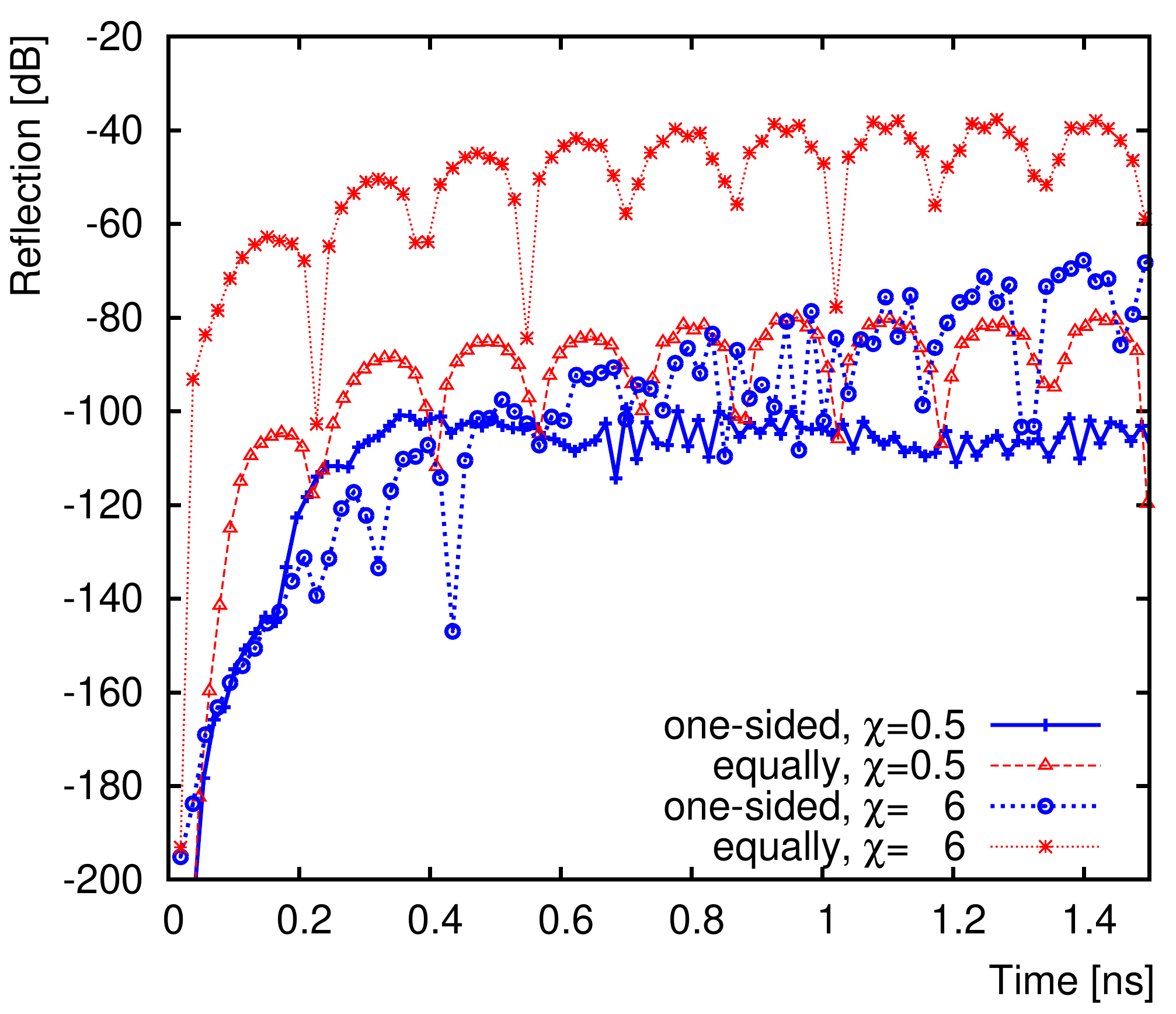}
    \end{tabular}
    \caption{Comparing the reflection at layers of perfectly matched media of the Divergence Preserving ADI and the original ADI scheme for $\chi = 6$ on the left. On the right side the reflections of four runs with the divergence preserving ADI scheme are compared. Two of these runs were carried out with the proposed $\lambda$'s (one-sided) while for the other two runs the conductivities were added half implicitly and half explicitly (equally). For both choices of the $\lambda$'s the simulation was carried out with two different lengths of the time step, i.e., $\chi=0.5$ and $\chi=6$.}
    \label{fig:PML_reflection}
  \end{center}
\end{figure}

For $\chi = 6$ the divergence preserving scheme is compared with the original ADI scheme. The Yee scheme cannot be used at such large time steps since it becomes unstable for $\chi > 1$. As one can observe in \figref{PML_reflection} (left) the two schemes produce the same result up to relative differences in the order of $10^{-10}$. The absolute difference, $\Delta_\text{ADI, DP-ADI} = R_\text{ADI} - R_\text{DP-ADI}$, is added to the plot. If one compares the reflections of the ADI scheme with $\chi = 6$ in \figref{PML_reflection} (left) with the Yee scheme with $\chi=1$ in \figref{PML_reflection_smithe} a considerable degradation of the performance of the PML can be observed. Nevertheless, the reflection is still below $-60~\text{dB}$, which is sufficient in many situations. In \figref{PML_reflection} (right) the divergence preserving ADI scheme with $\lambda_\xi^{s,1} = \lambda_\xi^{s,3} = 0.5$ (conductivities added half implicitly, half explicitly) is compared with the same scheme but the proposed $\lambda$'s for $\chi=0.5$ and $\chi=6$. As was already observed in~\cite{bib:wang2003} an improper choice of the weights $\lambda_\xi^{s,\,\tau}$ in \eqnref{adipml} leads to a comparable degradation as increasing the step size by a factor of $12$.
\section{Conclusions}
It has been shown how perfectly matched layers have to be formulated within the divergence preserving ADI scheme. Furthermore their performance has been validated by comparing the reflections for small time steps with results obtained with the Yee scheme. This comparison gives very good results.

For large time steps beyond the CFL limit the implementation has been compared with results using a non-divergence preserving ADI scheme. The divergence preserving scheme gives equally low reflections as the non-preserving scheme.

Perfectly matched layers combined with a divergence preserving ADI scheme allow to perform simulations of charged particles in (partially) open domains. With this implicit scheme, the time step in these simulations is not bounded by the CFL condition. In our implementation the computing time for one step on a single core was roughly three times as large for the ADI scheme as for the Yee scheme. So, if the time step in the ADI scheme is chosen three times larger than in the Yee scheme then the ADI scheme gives faster results with sufficient accuracy.

\bibliography{paper}
\bibliographystyle{elsarticle-num}

\end{document}